\documentclass[11pt]{article}

\usepackage{epsfig}
\usepackage{amssymb}
\usepackage{latexsym}

\newcommand{\Real}{\mathbb R}
\newcommand{\Proj}{\mathbb P}

\newcommand{\FigWidth}{4.5}

\def\ni{\noindent}

\newcommand {\stepSkip} {\vskip 0.5cm}
\newcommand {\itemSkip} {\vskip 0.2cm}
\newcommand {\quSkip} {\vskip 0.1cm}
\def\step#1{{{\bf $\underline{Step~#1: }$}}}
\newcommand{\nli}{\\ \noindent}
\newcommand{\npr}{\\ \indent}
\newcommand{\nqr}{\\ \indent}

\title{{Algebraic Curves in Parallel Coordinates } \\ { -- Avoiding the ``Over-Plotting'' Problem}}

\author{Zur Izhakian\thanks{e-mail: zzur@post.tau.ac.il, Department of Computer Science,
Faculty of Exact Sciences, Tel Aviv University, Ramat Aviv, 69978,
Tel Aviv,  Israel.}}

\begin{document}

\maketitle

\begin{abstract}
${\cal U}$ntil now the representation (i.e. plotting) of curve in
Parallel Coordinates is constructed from the point
$\leftrightarrow$ line duality. The result is a ``line-curve''
which is seen as the envelope of it's tangents. Usually this gives
an unclear image and is at the heart of the ``over-plotting''
problem; a barrier in the effective use of Parallel Coordinates.
This problem is overcome by a transformation which provides
directly the ``point-curve'' representation of a curve. Earlier
this was applied to conics and their generalizations. Here the
representation, also called dual, is extended to all planar
algebraic curves. Specifically, it is shown that the dual of an
algebraic curve of degree $n$ is an algebraic of degree at most
$n(n - 1)$ in the absence of singular points. The result that
conics map into conics follows as an easy special case. An
algorithm, based on algebraic geometry using resultants and
homogeneous polynomials, is obtained which constructs the dual
image of the curve. This approach has potential generalizations to
multi-dimensional algebraic surfaces and their approximation. The
``trade-off'' price then for obtaining {\em planar} representation
of multidimensional algebraic curves and hyper-surfaces is the
higher degree of the image's boundary which is also an algebraic
curve in $\|$-coords.
\end{abstract}

{\bf keywords}: Visualization, Parallel Coordinates, Algebraic
Dual Curves, Approximations of Algebraic Curves, Surfaces.

{\bf AMS} : 76M27

{\bf ACM} : F.2.1, I.1.1


\section{Parallel Coordinates}
\begin{figure}[!h]
\centering
\includegraphics[width=\FigWidth in]{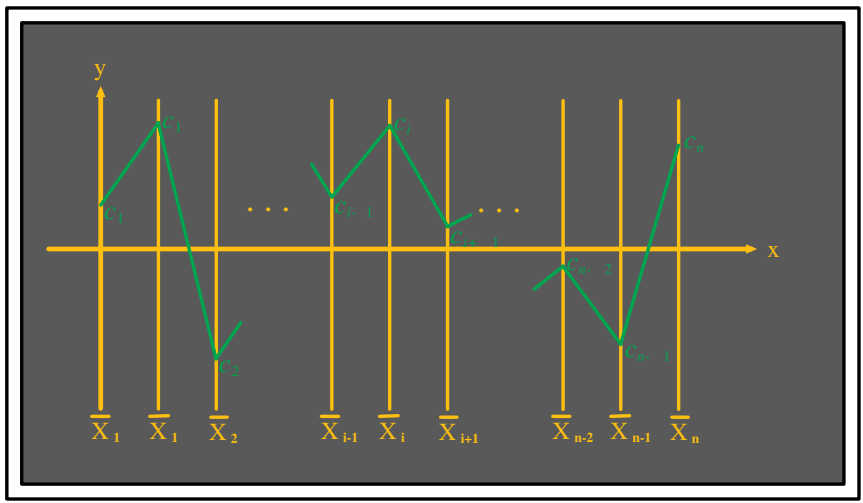}
\caption{\label{||-coords} A point $C = (c_{1},c_{2},\ldots,c_{n})
\in {\Real}^{n}$ is represented by the polygonal line $\bar{C}$
(consist of $n-1$ sections) with vertices at the $c_i$ values of
the $\bar{X}_{i}$ axis for $i=1,2,\ldots,n$.}
\end{figure}

\begin{figure}[!h]
\centering
\includegraphics[width=\FigWidth in]{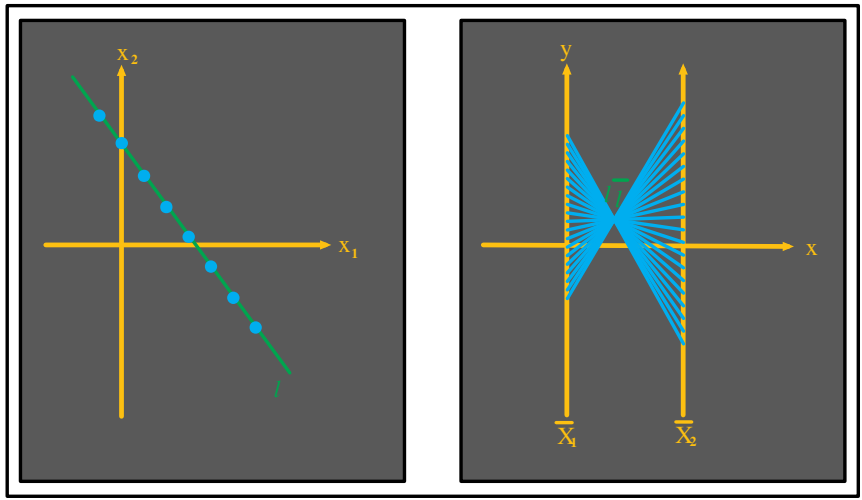}
\caption{\label{Dual} Fundamental $point\leftrightarrow line$
duality.}
\end{figure}

$\cal{O}$ver the years a methodology has been developed which
enables the visualization and recognition of multidimensional
objects \emph{ without loss of information}. It provides insight
into multivariate (equivalently multidimensional) problems and
lead to several applications. The approach of Parallel Coordinates
(abbr. $\|$-coords) \cite{inselberg85plane} is in the spirit of
Descartes, based on a coordinate system but differing in an
important way as shown in Fig \ref{||-coords}. On the Euclidean
plane ${\Real}^{2}$ (more precisely on the projective plane
${\Proj}^{2}$) with $xy$-Cartesian coordinates, $n$ copies of a
real line, labelled $\bar{X}_{1},\bar{X}_{2},\ldots, \bar{X}_{n}$,
are placed equidistant and perpendicular to the $x$-axis with
$\bar{X}_{1}$ and $y$ being coincident. These lines, which have
the same orientation as the $y$-axis, are the axes of the {\bf
Parallel Coordinate} system for the $n$-dimensional Euclidean
space ${\Real}^{n}$. A point $C = (c_{1},c_{2},\ldots,c_{n}) \in
{\Real}^{n}$ is represented by the polygonal line $\bar{C}$ having
vertices at the values $c_i$ on the $X_i$-axes. In this way a
one-to-one correspondence is established between a points in
${\Real}^{n}$ and planar polygonal lines with vertices on the
parallel axes. The polygonal line $\bar{C}$ contains the {\em
complete} lines and not just the segments between adjacent axes.
\npr
The restriction to ${\Real}^2$ provides that not only is a point
represented by a line, but that a line is represented by a point.
The points on a line are represented by a collection of lines
intersect at a single point as can be seen in Fig. \ref{Dual};
i.e. a ``pencil'' of lines in the language of Projective Geometry.
A fundamental $point \leftrightarrow line$ duality is induced
which is the cornerstone of deeper results in $\|$-coords. The
multidimensional generalizations for the representation of linear
$p$-flats in ${\Real}^n$ (i.e. planes of dimension $1 \leq p \leq
n - 1$), in terms of indexed points have been obtained. The proper
setting for dualities is the projective ${\Proj}^2$ rather than
the Euclidean plane ${\Real}^2$. A review of the mathematical
foundations is available in \cite{Insel99a}.
\begin{figure}[!h]
\centering
\includegraphics[width=\FigWidth in]{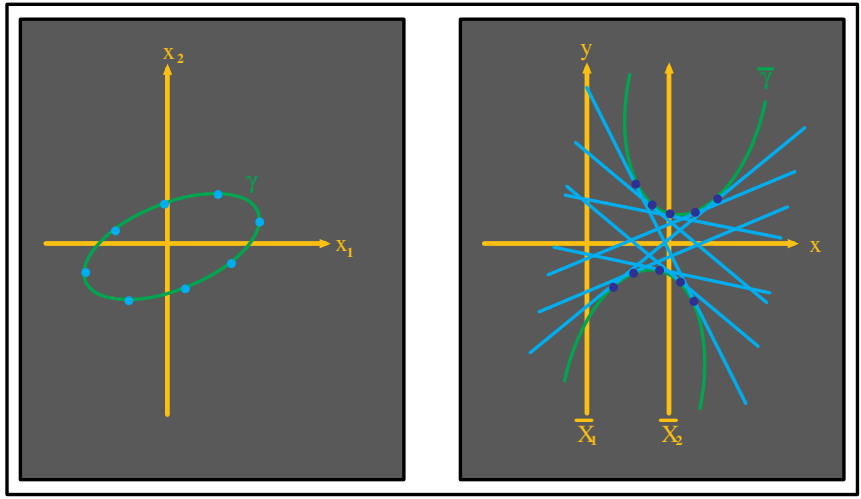}
\caption{\label{Envelope} Point-curve mapped into line-curve
(envelope of lines).}
\end{figure}
\npr
For non-linear, especially non-convex, objects the representation
is naturally more complex. In ${\Real}^{2}$ a point-curve, a curve
considered as collection of points, is transformed into a {\em
line-curve}; a curve prescribed by it's tangent lines as in Fig.
\ref{Envelope}. The line-curve's envelope, a point-curve,  is the
curve's image in $\|$-coords. In many cases this yields an image
that is difficult to discern. This point requires elaboration in
order to motivate and understand some of the development presented
here. As posed, the construction of a curve's image involves the
sequence of operations:
%
\npr $ \qquad point-curve \; \; \rightarrow \; \; line-curve  \;
\qquad \longmapsto $
\npr $ \qquad  point-curve \; (as \; the \; envelope \; of \;
line-curve). $
\smallskip
%
%
\npr
Unlike the example shown in Fig. \ref{Envelope}, where the
line-curve's image is clear, the plethora of overlapping lines
obscures parts of the resulting curve-line. This is a
manifestation of what is sometimes called ``over-plotting''
problem in $\|$-coords; the abundance of ``ink'' in the picture
covers many underlying patterns. Simply our eyes are not capable
of ``extracting'' the envelope of lots of overlapping lines. There
are also computational difficulties involved in the direct
computation of the image curve's envelope. In a way this is the
``heart'' of the ``over-plotting'' problem considered as a barrier
in the effective use of $\|$-coords. This problem can be overcome
by skipping the intermediate step and go to an equivalent
$$
point-curve \; \; \longmapsto \; \; point-curve
$$
transformation which provides {\bf directly} a clear image.
 as shown in Fig. \ref{Curve}.
\begin{figure}[!h]
\centering
\includegraphics[width=\FigWidth in]{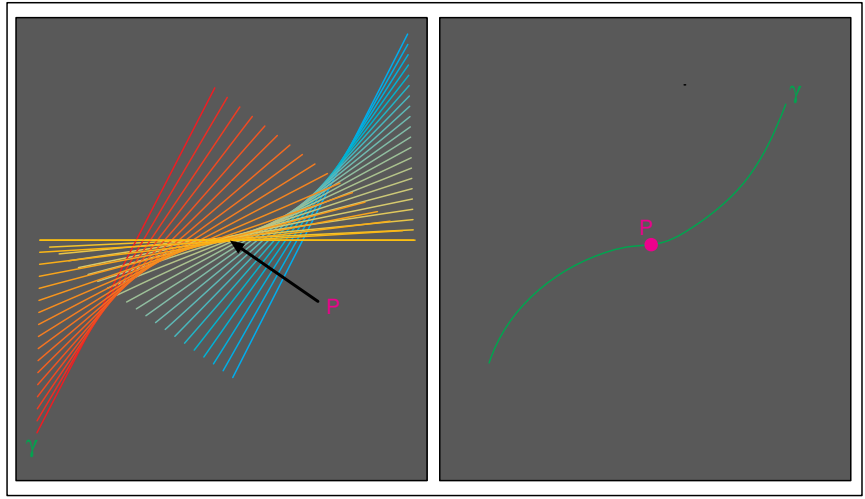}
\caption{\label{over-plot} In line-curve (on the left) the
tangents cover the image shown on the right as a point-curve.}
\end{figure}
\begin{figure}
\centering
\includegraphics[width=\FigWidth in]{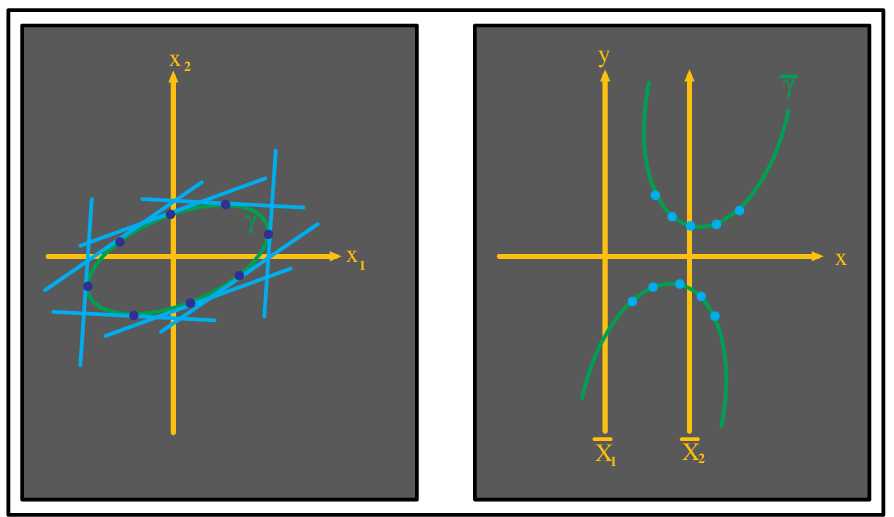}
\caption{\label{Curve} Point-curve mapped into point-curve.}
\end{figure}
%
%
This point-to-point mapping (Inselberg Transformation
\cite{inselberg85plane} ) preserves the curve's continuity
properties. The idea is to use the $\rightarrow$ part of the
duality and map the {\em tangents} of the curve into points as
illustrated in Fig. \ref{Curve}. Therefore, a curve is represented
by a point-curve (the ``dual curve'') in the $\|$-coords plane.
For a point-curve, $\gamma$, defined \emph{implicitly} by
\begin{equation}\label{eq:ptc-implicit}
  \gamma \,: \, f (x_1, x_2) = 0.
\end{equation}
\ni the $x, y$ coordinates of the point-curve image (i.e. dual)
$\bar{\gamma}$ are given by
\begin{equation}\label{eq:F2f}
\left\{
\begin{array}{rrl}
x&=& \frac{{\partial f}/{\partial x_2}}{({\partial f}/
            {\partial x_1} + {\partial f}/{\partial x_2})} \; , \\[2mm]
y&=& \frac{(x_1 {\partial f}/{\partial x_1} + x_2{\partial f}/
{\partial x_2})}{({\partial f}/{\partial x_1} + {\partial
f}/{\partial x_2})}.
\end{array}
\right.
\end{equation}
%
%
\npr
It was shown by B. Dimsdale \cite{dimsdale84conic} and generalized
in \cite{inselberg85plane}, using this transformation that conics
are mapped into conics in 6 different ways \cite{dimsdale84conic}.
Here we develop the extension of the dual image for the family of
general algebraic curves. This family of curves is highly
significant in many implementations and applications, since these
curves can easily and uniquely be reconstructed from a finite
collection of their points (for instance using simple
interpolation methods \cite{burden89numerical},
\cite{walter97numerical}). Approximations of such curves can
simply be obtained using similar methods.
\npr
As will be seen, the dual of an algebraic curve in general has
degree higher than the original curve. There are some
fringe-benefits, illustrated later, where ``special-points'' such
as self-intersections or inflection-points which are conveniently
transformed. But we do not want to get ahead of ourselves. The key
reason for this effort is to pave the way for the representation
of algebraic curves and more general hyper-surfaces, as well as
their approximations, in terms of planar regions {\em without
losing information}. To pursue this goal then we need firstly to
study the image of curves starting with general algebraic curves.
%
\section{Transforms of Algebraic Curves}
\subsection{The Idea Leading to the Algorithm} ${\cal I}$n order to
represent non-linear relations in $\|$-coords it is essential to
extend the representational results first to algebraic curves;
those described by either, explicitly or implicitly by irreducible
polynomials of arbitrary degree. The direct application of eq.
(\ref{eq:F2f}) turns out to be difficult even for degree 3. There
is a splendid way to solve this problem using some ideas and tools
from Algebraic Geometry. These involve properties of homogeneous
polynomials and Resultant which are explained during the
development of the method. For extensive treatments the reader is
referred to \cite{cox97ideals}, \cite{harris92algebraicm},
\cite{hodge52methods} and \cite{walker78algebraic}.
\npr
Starting with an algebraic curve $\gamma$ defined by an
irreducible polynomial $f(x_1,x_2) = 0$ in ${\Real}^2$ its image
in $\|$-coords is sought. A number of preparatory steps smooth the
way for the easier application of the transformations in eq.
(\ref{eq:F2f}). First the curve $\gamma \subset {\Real}^{2}$ is
``raised'' to a surface embedded in the projective space
${\Proj}^{2}$ and only then the corresponding mapping into
$\|$-coords is applied.
\begin{description}
\stepSkip
\item[\step{1}]
There exists a one-to-one correspondence (preserving the
reducibility of polynomials) between any polynomial
$f(x_{1},x_{2}) = 0 $ of degree $n$ and a homogenous polynomial in
the projective plane ${\Proj}^{2}$, which is obtained by using
homogeneous coordinates. Specifically,
\begin{itemize}
  \item replace $x_k$ by $\frac{x_k}{x_3}$ for $k = 1, 2$,
  \item multiply the whole polynomial by ${x_3}^n$ and simplify.
\end{itemize}
\noindent These multiplications are, of course, allowable for $x_3
\not = 0$. The result is a \emph{homogeneous polynomial} (which is
also irreducible if $f = 0$ is) with each term having degree $n$.
To wit,
$$  f(x_{1},x_{2})=\sum_{i+j=0}^{n}a_{ij}x_{1}^{i}x_{2}^{j} \hskip
5pt \rightarrow \hskip 15pt \qquad $$
\begin{equation}\label{eq:nt}
F(x_{1},x_{2},x_{3})=\sum_{i+j=0}^{n}a_{ij}x_{1}^{i}x_{2}^{j}x_{3}^{n-i-j},
\end{equation}
\noindent which describes a surface in ${\Proj}^2$ where the
original polynomial curve is embedded as:
\begin{equation}\label{eq:emb}
  f(x_{1},x_{2}) = F(x_{1},x_{2}, 1 ).
\end{equation}
\stepSkip

\begin{figure}[!h]
\centering
\includegraphics[width=\FigWidth in ]{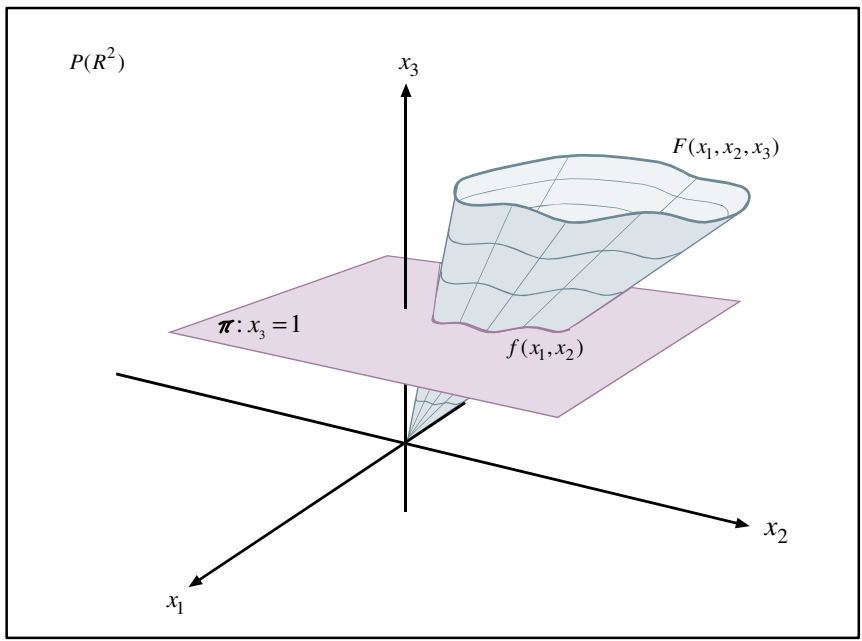}
\caption{\label{Cone} Cone $F(x_1, x_2, x_3) = 0$ generated by
embedding the curve $f(x_1, x_2) = 0$ in ${\Proj}^2$.}
\end{figure}
%

\item[\step{2}]
The \emph{Gradient} of $F$ is found and denoted by :
$$ \nabla F (x_{1},x_{2},x_{3}) = (\frac{\partial
F}{\partial x_1} \; , \; \frac{\partial F}{\partial x_2} \; , \;
\frac{\partial F}{\partial x_3}) = (\eta, \xi, \psi ). $$ \ni The
three derivatives provide the direction numbers of the normal to
the tangent plane at the point $({x}_1, {x}_2, {x}_3)$ . It is a
fundamental property of homogeneous polynomials that
\begin{equation}\label{eq:homder}
  x_1 \frac{\partial F}{\partial x_1}  + x_2 \frac{\partial
F}{\partial x_2}  + x_3 \frac{\partial F}{\partial x_3} = n F
\end{equation}
\noindent where $n$ is the degree of $F$. In our case $ F = 0$ and
hence the equation of the tangent planes is
\begin{equation}\label{eq:tp}
  {\eta} x_1 + {\xi} x_2 + {\psi} x_3 = 0.
\end{equation}
\noindent Since the tangent plane at \emph{any} point of the
surface goes through the origin it is clear that the surface $F$
is a {\bf cone} with apex at the origin as shown in Fig.
\ref{Cone}. \stepSkip
\item[\step{3}]
Substituting for $x_3$ from eq. (\ref{eq:tp}) in
\begin{equation}
 \label{eq:Fsub}
 F(x_1, x_2, -(\frac{{\eta} x_1 + {\xi} x_2}{\psi})) = 0.
\end{equation}
 \noindent provides the intersection of the tangent plane
with the cone which is a whole line as shown in Fig.
\ref{gen-cone}. Each one of these lines, of course, goes through
the origin and therefore it can be described by any one other of
its points and in particular by $(x_1, x_2, 1)$. Simplifying the
homogeneous coordinates by,
$$(x_1, x_2,
-(\frac{{\eta} x_1 + {\xi} x_2}{\psi}) = ({\psi}x_1, {\psi}x_2,
-({\eta} x_1 + {\xi} x_2)) $$
\noindent results in
\begin{equation}\label{eq:glines}
  F({\psi}x_1,{\psi}x_2, -({\eta} x_1 + {\xi} x_2)) = 0.
\end{equation}
\ni By the way, this is also a homogeneous polynomial in the
\emph{five} variables appearing in its argument.
\begin{figure}[!h]
\centering
\includegraphics[width=\FigWidth in]{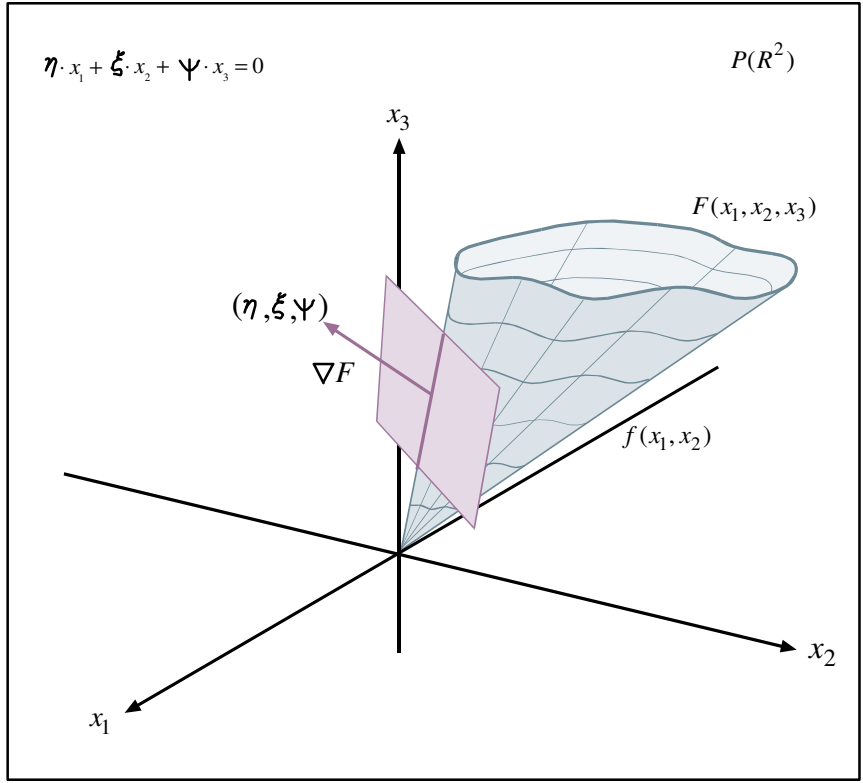}
\caption{\label{gen-cone} The intersection of the cone
$F(x_{1},x_{2},x_{3})= 0$ with any of it's tangent planes is a
whole line. Along such a line the direction numbers $(\eta, \xi,
\psi )$ are constant and unique.}
\end{figure}
It is helpful to understand the underlying geometry. Eq.
(\ref{eq:glines}) can be considered as specifying a family of
lines through the origin $(0, 0, 0)$ and through each point $(x_1,
x_2, 1)$ along the curve given by eq. (\ref{eq:emb}).
Alternatively, the cone can be considered as being generated by a
line, the \emph{generating line} -- see Fig. \ref{gen-line},
pivoted at the origin and moving continuously along each point
$(x_1, x_2, 1)$ of the curve described by eq. (\ref{eq:emb}). On
each one of these lines the direction numbers $ (\eta, \xi, \psi)$
are unique and constant as shown in Fig. \ref{gen-line}; i.e.
there is a one-to-one correspondence:
$$ (x_1, x_2, 1) \leftrightarrow (\eta, \xi, \psi).$$
\ni Eq. (\ref{eq:Fsub}), and hence it's rephrasing eq.
(\ref{eq:glines}), contains two equivalent descriptions of the
cone, i.e. in terms of the coordinates $(x_1, x_2, 1)$ and also
the direction numbers $(\eta, \xi, \psi)$. Hence one can be
eliminated and as it turns out it is best to eliminate the $x_i \,
, \, i = 1, 2$ something which is very conveniently done by means
of the \emph{Resultant}. The resultant $R(F,G)$ of two homogenous
polynomials \quSkip \qquad $F(x_1, x_2) = \sum_{i =0}^n
a_{i}{x_1}^i {x_2}^{n-i} = 0 $ and \nqr \qquad $G(x_1,x_2) =
\sum_{j =0}^m b_{j}{x_1}^j {x_2}^{m-j} = 0 $ \quSkip is the
polynomial obtained from the determinant of their coefficients
matrix:
\quSkip
$R(F,G)=$
$$
 Det \left( \matrix{
 a_{0} & a_{1} & a_{2} & \cdots & a_{n} &  &  & 0 \cr
 & a_{0} & a_{1} &  &  & a_{n} &  &  \cr
 &  & \ddots &  &  &  &\ddots &  \cr
 &  &  & a_{0} &  & \cdots &  & a_{n} \cr
 b_{0} & b_{1} & b_{2} & \cdots & b_{m} &  &  &  \cr
 b_{0} & b_{2} &  &  & b_{m} &  &  \cr
 &  & \ddots &  &  &  & \ddots &  \cr
 0 &  &  & b_{0} &  & \cdots &  & b_{m} \cr
} \right),$$
 \noindent where the empty spaces are filled by zeros. There are
$m$ lines of $a_i$ and $n$ lines of $b_j$. When $F$ and $G$ are
both irreducible so is their resultant. Due to the homogeneity of
$F$ and $G$,
$$ R(F, G) = 0 \Leftrightarrow F = 0 \; , \; G = 0 \; . $$
\ni Let us rewrite eq. (\ref{eq:glines}) as
\begin{equation}\label{eq:Fne}
  F_{(\eta, \xi, \psi)}(x_1, x_2) = \sum_{i =0}^n a'_{i}(\eta,
\xi, \psi ){x_1}^i {x_2}^{n-i} = 0 \; .
\end{equation}
\ni Appealing again to the homogeneity of $F$ to obtain the
relation
$$ x_1 \frac{\partial F}{\partial x_1}  +
   x_2 \frac{\partial F}{\partial x_2} = n F $$
\ni which has already been mentioned earlier. Therefore
$$\frac{\partial F}{\partial x_1} = 0 \; \; , \; \; \frac{\partial
F}{\partial x_2} = 0 \Rightarrow \; \; \; F = 0 $$
\noindent and from the property of the resultant of homogeneous
polynomials mentioned above
$$ R (\frac{\partial F}{\partial x_1} ,  \frac{\partial F}{\partial
x_2}) = 0 \Leftrightarrow \frac{\partial F}{\partial x_1} = 0 \;
\; , \; \; \frac{\partial F}{\partial x_2} = 0. $$
Altogether then
\begin{equation}\label{eq:FR}
 R (\frac{\partial F}{\partial x_1} ,  \frac{\partial F}{\partial
x_2}) = R(\eta, \xi, \psi ) = 0
\end{equation}
\ni has finite degree since $F$ and $R$ are polynomials. It was
shown that
$$ R = 0 \Leftrightarrow F = 0 $$
\ni so not only their zero sets agree, which ensures that they are
equivalent polynomials, but they also agree at an infinite number
of points. \noindent Hence, $R = 0$ provides the description of
the cone $F = 0$ in terms of the direction numbers $(\eta, \xi,
\psi )$ and is also a homogeneous polynomial. The degree of
$\frac{\partial F}{\partial x_i}$ is $n-1$. This and the structure
of the resultant with the four triangular zero portions results in
$R$ being a polynomial of degree at most $n(n-1)$.
\begin{figure}[!h]
\centering
\includegraphics[width=\FigWidth in]{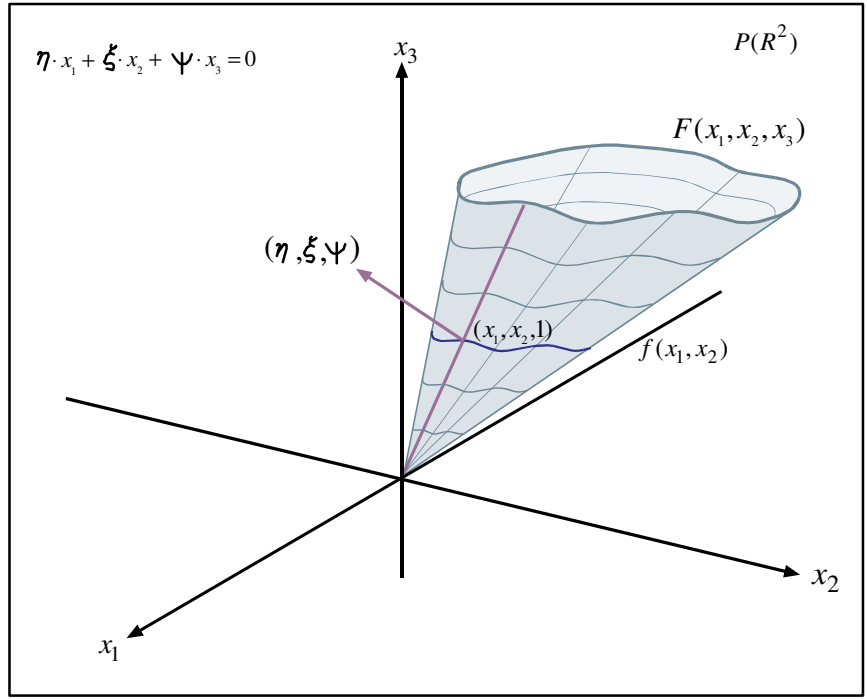}
\caption{\label{gen-line} This shows the \emph{generating line} of
the cone (which is found from the intersection of the tangent
planes with the cone) together with the gradient vector.}
\end{figure}
\stepSkip
\item[\step{4}]
Since we are interested in the solutions of $ R = 0$ the
multiplier in powers of $\psi$ of $R$, if there is one, can be
safely neglected due to the homogeneity. \stepSkip
\item[\step{5}]
Only in this, the final step, the transformation of the curve from
the $x_1 x_2$-plane to the $xy$-plane with parallel coordinates is
performed. It is done using equations (\ref{eq:F2f}) rewritten, in
view of eq. (\ref{eq:tp}) as
\begin{equation}\label{eq:F2final}
  x = \frac {\xi}{\eta + \xi} \; \; , \; \; y = - \frac{\psi}{\eta
+ \xi} \; .
\end{equation}
\ni It is important to notice that \emph{both} this and eq.
(\ref{eq:FR}) \emph{involve only the derivatives} $\eta, \xi,
\psi$. Let $c = {\eta + \xi}$, so that $\xi = c x$, $\eta = c - cx
= c(1-x)$ and $\psi = - cy$. Substitution provides the transform
of the original curve $f(x_1, x_2) = 0$ :
$$ R(\eta, \xi, \psi) = R ( c(1-x), cx , - cy)
= $$
$$c ^n R ((1-x), x, -y) = 0 \; \; \Rightarrow $$
$$R ((1-x), x,-y) = 0 \; ,$$
\ni when $c \not = 0$. In the previous step it was pointed out
that this polynomial has degree at most $n(n-1)$, the actual
degree obtained depends on the presence of singular points in the
original algebraic curve $f = 0$.
\end{description}
\subsection{Algorithm} ${\cal H}$ere the process involved is
presented compactly as an algorithm \cite{zur:01:ms} whose input
is an algebraic curve $\gamma : f(x_{1},x_{2}) = 0$ and the output
is the polynomial which describes $\bar{\gamma}$, the curve's
image in $\|$-coords. To emphasize, the algorithm applies to
implicit or explicit polynomials of any degree and curves with or
without singular points. The formal description of the algorithm
is followed by examples which clarify the various stages and their
nuances.
\npr
For a given irreducible polynomial equation (otherwise apply the
algorithm for each of it's component separately)
  $f(x_{1},x_{2}) = 0$.
\begin{enumerate}
  \item Convert to homogeneous coordinates to obtain the transformation of $f$, a homogenous polynomial $F(x_{1}, x_{2}, x_{3})=
0$.
\itemSkip
  \item Substitute
\npr \qquad
  $x_{1} \rightarrow \psi x_1 $, $\; x_{2} \rightarrow \
\psi x_2 \;$ and \npr \qquad $\; x_{3} \rightarrow -(\eta x_1+\xi
x_2 )$.
\itemSkip
  \item Find the resultant of the two derivatives $F_{x_1}$ and
  $F_{x_2}$.
\itemSkip
  \item Cancel the multiplier in a power of $\psi$ of the resultant $R$
and denote the result by $R'$.
\itemSkip
  \item The output is obtained by the substitution
\npr \qquad
  $\eta =1-x$, $\; \xi = x$, $ \; \psi = -y \; $ in $ \; R'$.
\end{enumerate}
\section{Examples of Algebraic Curves and their Transforms}
\subsection{Conic Transforms} 
\begin{figure}[!h]
\centering
\includegraphics[width=\FigWidth in]{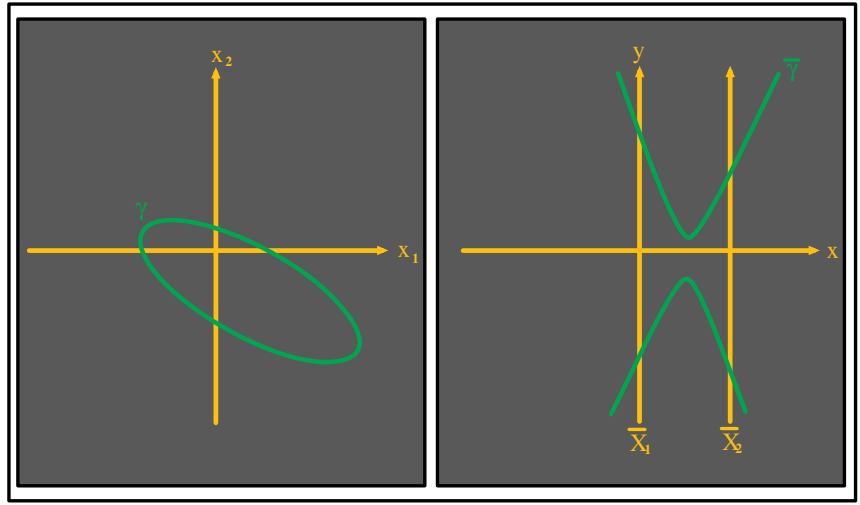}
\caption{\label{fig2to2} An ellipse is mapped into hyperbola.}
\end{figure}

${\cal T}$he algorithm is
illustrated with some examples starting with the conics. \nli
$f(x_{1},x_{2)}= \left( \matrix{
                 x_{1} & x_{2} & 1 \cr}
                 \right)
\left( \matrix{
        A_{1} & A_{4} & A_{5} \cr
        A_{4} & A_{2} & A_{6} \cr
        A_{5} & A_{6} &A_{3} \cr }
 \right)
 \left( \matrix{
 x_{1} \cr
 x_{2} \cr
 1 \cr }
\right) =$ \nli \vskip 0.1in
$A_{1}x_{1}^{2}+2A_{4}x_{1}x_{2}+2A_{5}x_{1}+A_{2}x_{2}^{2}+2A_{6}x_{2}+A_{3}=0.$
\vskip 0.1in \ni Applying the algorithm in the sequence given
above: \stepSkip
\begin{description}
\item[\step{1}]
The homogeneous polynomial obtained from $f$ by using homogeneous
coordinates is
\smallskip
\nqr $\qquad F(x_{1},x_{2},x_{3})=$
\nqr $\qquad \qquad
A_{1}x_{1}^{2}+2A_{4}x_{1}x_{2}+2A_{5}x_{1}x_{3} +$
\nqr $\qquad \qquad
A_{2}x_{2}^{2}+2A_{6}x_{2}x_{3}+A_{3}x_{3}^{2}=0.$
\stepSkip
\item[\step{2}]
Substituting \hskip 2mm
\smallskip
\nqr $\qquad x_{1} \rightarrow \psi x_1,
  \hskip 2mm x_{2} \rightarrow \psi x_2,\hskip 2mm  x_{3} \rightarrow -(\eta
x_1 +\xi x_2 ) \hskip 2mm $ \nli yields
\smallskip
\nqr $\qquad F(\psi x_1,\psi x_2,-(\eta x_1+\xi x_2))=$
\smallskip
\nqr $\qquad \qquad  (-2A_{5}\psi \eta +A_{1}\psi ^{2}+A_{3}\eta
^{2})+ {x_1}^{2}$
\nqr $\qquad \qquad 2( -A_{5}\psi \xi +A_{3}\xi \eta +A_{4}\psi
^{2}-A_{6}\psi \eta ) x_1 x_2+$
\nqr $\qquad \qquad (A_{2}\psi ^{2}-2A_{6}\psi \xi +A_{3}\xi
^{2}){x_2}^{2}=$
\smallskip
\nqr $\qquad \qquad c_1 {x_1}^{2} + 2 c_2 x_1 x_2 + 2 c_3
{x_2}^{2} = 0, $
\smallskip
\\
\qquad where $ c_i = c_i(\eta , \xi , \psi)$, for $i = 1,2,3$.
\stepSkip
\item[\step{3}]
Calculate the two derivatives of $F$ and their resultant:
\smallskip
\nqr $\qquad \frac{\partial F}{\partial x_1}=2 c_1 x_1 + 2 c_2
x_2,$ 
\smallskip
\nqr $\qquad \frac{\partial F}{\partial x_2}=2 c_2 x_1 +2 c_3
x_2,$
\smallskip
\nqr $\qquad R(\frac{\partial F}{\partial x_1},\frac{\partial F}
{\partial x_2} ) =  Det \left( \matrix{2 c_1 & 2 c_2 \cr 2 c_2 & 2
c_3 \cr} \right)=$
\smallskip
\nqr $\qquad {R}(\eta ,\xi ,\psi ) =  $
\smallskip
\nqr $\qquad \qquad  4\psi ^{2} [ \left(
A_{3}A_{2}-A_{6}^{2}\right) \eta ^{2}+ \left(
A_{1}A_{3}-A_{5}^{2}\right) \xi ^{2}+$
\nqr $\qquad \qquad \left( A_{1}A_{2}-A_{4}^{2}\right) \psi
^{2}+2(A_{5}A_{6}-A_{3}A_{4})\eta \xi +$
\nqr $ \qquad \qquad  2(A_{5}A_{4}-A_{1}A_{6})\xi \psi
+2(A_{4}A_{6}- 2A_{5}A_{2})\eta \psi.$ \stepSkip
\item[\step{4}]
Retain the resultant's component which is multiplied by a power of
$\psi$ and let
\smallskip
\nqr $\qquad {R'}(\eta ,\xi ,\psi )=\frac{{R} (\eta ,\xi ,\psi
)}{4\psi ^{2}}.$
\stepSkip
\item[\step{5}]
The dual of $ f(x_1, x_2) = 0$ is then given in matrix form by
\nqr $\qquad {R'}(1 - x, x, -y) = $ \nqr $\qquad  \qquad  \left(
\matrix{ x & y & 1 } \right) \left( \matrix{
            a_1 & a_4 & a_5 \cr
            a_4 & a_2 & a_6 \cr
            a_5 & a_6 & a_3 \cr
         } \right)
\left( \matrix{
    x \cr
    y \cr
    1 \cr
} \right), $
\vskip 0.1in where the individual $a_i$ are :
\smallskip
\nqr $ \qquad a_{1} = A_3 (A_1 + A_2 + 2A_4) - (A_5 + A_6)^2,$
\smallskip
\nqr $ \qquad  a_{2} = A_1 A_2 - {A_4}^2,$
\smallskip
\nqr $ \qquad  a_{3} = A_2 A_3 - {A_6}^2,$
\smallskip
\nqr $ \qquad a_{4} = A_6(A_1 +A_4) - A_5(A_2 +A_4),$
\smallskip
\nqr $ \qquad a_{5} = {A_6}^2 + A_5A_6 - A_3(A_2 +A_4),$
\smallskip
\nqr $ \qquad a_{6} = A_2 A_5 - A_4 A_6 .$
\smallskip
\end{description}
The result is illustrated for the ellipse in fig \ref{fig2to2} and
can also be contrasted to the earlier ways for obtaining the
transformation
\[conic \quad \leftrightarrow \quad conic. \]
%
\subsection{Algebraic Curves of Degree Higher than Two} ${\cal
N}$ext the algorithm is applied to the algebraic curve of 3rd
degree
$$ f(x_{1},x_{2})=x_{1}^{3}-x_{1}^{2}-x_{2}^{2}+x_{2}-1=0. $$
\begin{figure}[!h]
\centering
\includegraphics[width=\FigWidth in]{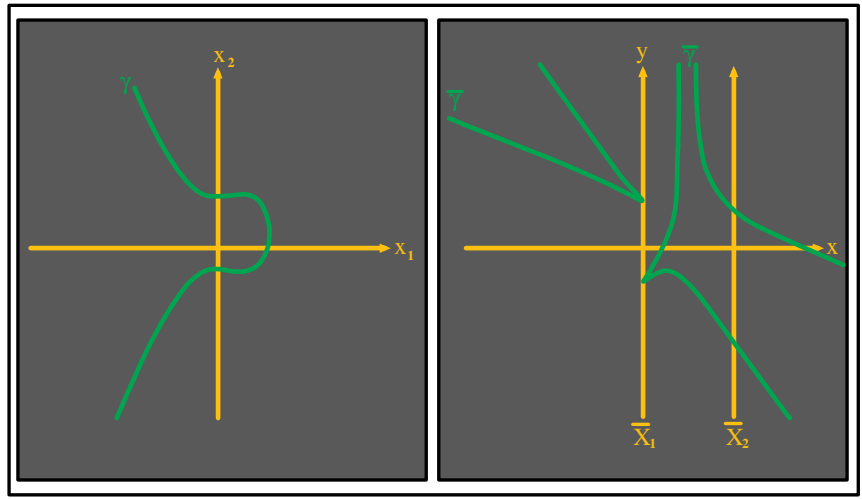}
\caption{\label{fig3to6} The cubic curve
$x_{1}^{3}-x_{1}^{2}-x_{2}^{2}+x_{2}-1= 0$ (on the left) is mapped
into a six degree curve (on the right). Notice also an instance of
the duality {\em inflection-point} $\leftrightarrow$ {\em cusp}
[4].
}
\end{figure}
\begin{figure}[!h]
\centering
\includegraphics[width=\FigWidth in]{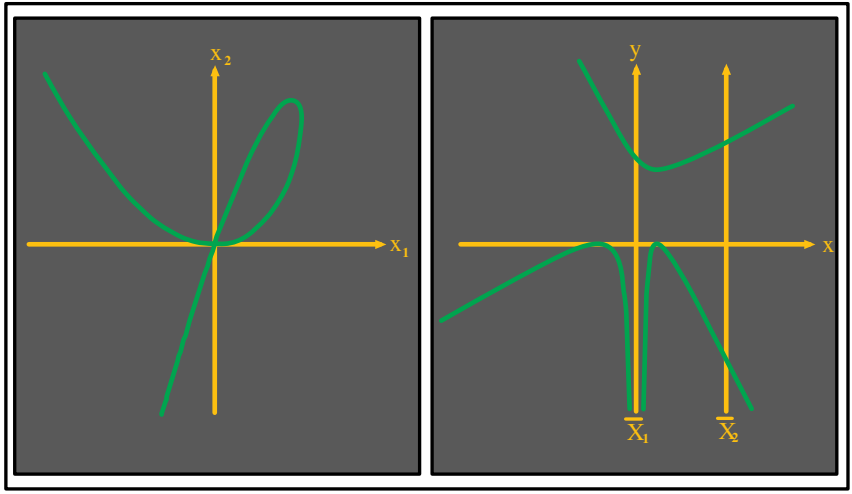}
\caption{\label{fig3to4} The self intersection point in the cure
$x_{1}^{3}+x_{2}^{2}-3x_{1}x_{2}=0$ (on the left), disappears in
the image curve
$27y^{2}-54y^{2}x+27y^{2}x^{2}-108y+270yx-198yx^{2}+40yx^{3}+108x^{3}-36x^{4}-81x^{2}=0$
 (on the right)
.}
\end{figure}
\stepSkip
\begin{description}
\item[\step{1}]
The homogeneous polynomial obtained from $f$ by using homogeneous
coordinates is
\smallskip
\nqr
 $ \qquad F(x_{1},x_{2},x_{3})=x_{1}^{3}-x_{1}^{2}x_{3}-x_{2}^{2}x_{3}
  + x_{2}x_{3}^{2}-x_{3}^{3}=0 . $
\stepSkip
\item[\step{2}]
Substituting \hskip 2mm
\smallskip
\nqr $ \qquad x_{1} \rightarrow \psi x_1,
  \hskip 2mm x_{2} \rightarrow \psi x_2,\hskip 2mm  x_{3} \rightarrow -(\eta
x_1 +\xi x_2 ) \hskip 2mm $
\smallskip
\nli yields
\smallskip
\nqr $\qquad F(\psi x_1,\psi x_2,-(\eta x_1+\xi x_2))=$
\nqr $\qquad \qquad( \psi ^{3}+\psi ^{2}\eta -\eta ^{3}) {x_1}^{3}
+ $
\nqr $\qquad \qquad (-3\xi \eta ^{2}+ \psi \eta ^{2}+ \psi ^{2}\xi
) {x_1}^{2}x_2  +$
\nqr $\qquad \qquad ( 2\psi \xi \eta +\psi ^{2}\eta -3\xi ^{2}\eta
)x_1 {x_2}^{2} + $
\nqr $\qquad \qquad (-\xi ^{3}+\psi \xi ^{2}+\psi ^{2}\xi
){x_2}^{3}=$
\smallskip
\nqr $\qquad \qquad c_1 {x_1}^{3} + c_2 {x_1}^{2}{x_2} + c_3
{x_1}{x_2}^{2} + c_4 {x_2}^{3}=0,$
\smallskip
\nli where the $c_i = c_i(\eta , \xi , \psi)$, for $i =1,...,3$.
\stepSkip
\item[\step{3}]
Calculate the resultant of the two derivatives of $F$:
\smallskip
\nqr $\qquad \frac{\partial F}{\partial x_1}=3 c_1 {x_1}^{2}+2 c_2
x_1 x_2 + c_3 {x_2}^{2},$
\smallskip
\nqr $\qquad \frac{\partial F}{\partial x_2}= c_2 {x_1}^{2}+2 c_3
x_1 x_2 + 3 c_4{x_2}^{2}, $
\smallskip
\nqr $\qquad R(\frac{\partial F}{\partial x_1},
  \frac{\partial F}{\partial x_2} )=
 Det\left(
\matrix{3 c_1 & 2 c_2 & c_3 & 0 \cr
        0 & 3 c_1 & 2 c_2 & c_3 \cr
        c_2 & 2 c_3 & 3 c_4 & 0 \cr
        0 & c_2 & 2 c_3 & 3 c_4 \cr
}\right) = $
\smallskip
\nqr $\qquad {R}(\eta ,\xi ,\psi )= -3\psi ^{6} {R'}(\eta , \xi ,
\psi), $
\smallskip
\nli \ni where ${R'}$ is polynomial of degree 6.
\stepSkip
\item[\step{4}]
Discarding the resultant's factor $(- {\psi}^6)$ i.e. let
\smallskip
\nqr $\qquad {R'}(\eta ,\xi ,\psi )=\frac{{R }(\eta ,\xi ,\pi
)}{-3\psi ^{6}} .$ \stepSkip
\item[\step{5}]
Finally, substituting
\smallskip
\nqr $\qquad \eta =1-x \; ,\; \xi =x \; , \; \psi =-y$
\smallskip
\nli results in the 6th-degree curve :
\smallskip
\nqr $\qquad  {R'}(1-x,x,-y)=$
\smallskip
\nqr $\qquad \qquad
-23+292y^{2}x^{2}-422x^{2}+326yx^{3}-146y^{2}x$
\nqr $\qquad \qquad +
610x^{3}+23y^{2}-27y^{4}x^{2}+54y^{4}x-27y^{4}$
\nqr $\qquad \qquad -22yx^{5}-244y^{2}x^{3}-66yx^{4}-420yx^{2}$
\nqr $\qquad \qquad
-126y^{3}x+232yx+214x^{5}-499x^{4}+90y^{3}x^{2}$
\nqr $\qquad \qquad +54y^{3}+156x-50y-31x^{6}+71y^{2}x^{4}$
\nqr $\qquad \qquad -14y^{3}x^{3}.$
\smallskip
\end{description}
\ni The source and image curve are shown in Fig. \ref{fig3to6}.
\noindent In this case $n=3$ and the maximum degree $n(n-1) = 6$
is attained. Notice that this covers as special cases :
\nli -- the point $\leftrightarrow$ line duality where points
(with $ n = 0$) are mapped into lines (with $n = 1$) and
vice-versa, and
\nli -- the conics with $n = 2$.
\nli
The next two examples show that the image curve may have degree
less than $n(n-1)$. Apparently the maximal degree is attained by
the image curve in the absence of singularities in the function or
it's derivatives. The full conditions relating the image's degree
to less then the maximal are not known at this stage. Two such
examples are shown in the subsequent figures, Fig. \ref{fig3to4}
and Fig. \ref{fig3to3}.
%
\begin{figure}[!h]
\centering
\includegraphics[width=\FigWidth in]{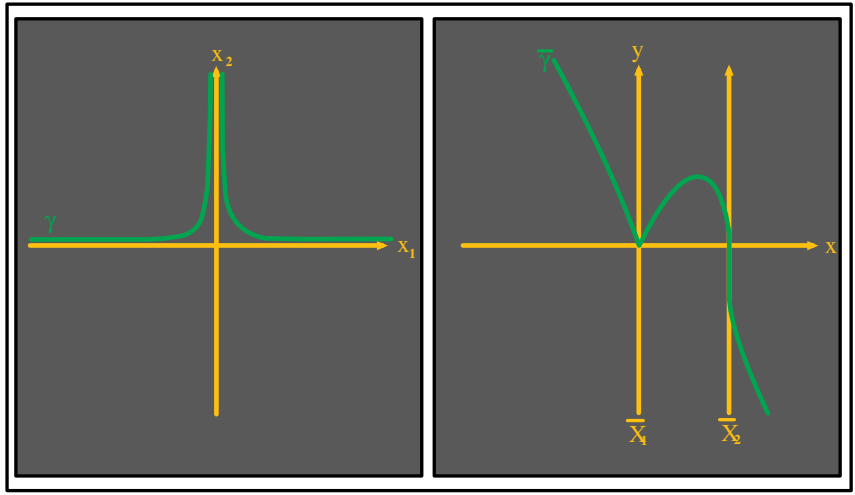}
\caption{\label{fig3to3} $x_{1}^{2}x_{2}-1=0 \; \; \longmapsto \;
\; 4y^{3}+27x^{3}-27x^{2}=0$.}
\end{figure}
%
\section{Conclusions} $\cal{C}$onsider the class $\cal{S}$ of
hyper-surfaces in ${\Real}^n$ which are the envelopes of their
tangent hyper-planes. As mentioned earlier \cite{Insel99a}
hyper-planes can be represented in $\|$-coords by $n - 1$ indexed
points. Hence for a hyper-surface $\sigma \in \cal{S}$ each point
$P \in \sigma$ maps into $ n - 1 $ indexed planar points. From
this it follows that the hyper-surface $\sigma$ can be mapped
(i.e. represented) by $n - 1$ indexed planar regions
${\bar{\sigma}}_i$ composed of these points. Restricted classes of
hyper-surfaces have been represented in this way and it turns out
that the ${\bar{\sigma}}_i$ reveal non-trivial properties of the
corresponding hyper-surface $\sigma$. It has already been proved
that for any dimension {\em Quadrics} (algebraic surfaces of
degree 2) are mapped into planar regions whose boundaries are {\em
conics} \cite{zur:04:wscg}. This also includes non-convex surfaces
like the ``saddle''. There are strong evidence supporting the
conjecture that algebraic surfaces in general map into planar
regions bounded by algebraic curves. This, besides their
significance on their own right, is one of the reasons for
studying the representation of algebraic curves. In
\cite{matskewich99approx} families of approximate planes and flats
are beautifully and usefully represented in $\|$-coords. Our
results cast the foundations not only for the representation of
hyper-surfaces in the class $\cal(S)$, but also their
approximations in terms of {\bf planar curved regions}.
\section*{Acknowledgments}
$\cal{T}$he author would like to thank to Prof. Alfred Inselberg
for his grate help and his kindly support. The author acknowledge
and is grateful for the use of the symbolic manipulation program
{\em Singular} developed by the Algebraic Geometry Group,
Department of Mathematics, University of Kaiserslautern, Germnay.
\nocite{*}
\nocite{ex1,ex2}
\bibliographystyle{latex8}
\bibliography{dfz}

%

\end{document}